\documentclass[amsmath,amssymb]{PoS}
\usepackage{amsfonts}
\usepackage{bm}
\usepackage{dcolumn}
\title{Lattice study of the leptonic decay constant of the pion and
  its excitations}

\ShortTitle{Lattice study of the leptonic decay constant of the pion
  and its excitations}

\author{\speaker{Ekaterina V.\ Mastropas}%
\\
        College of William and Mary, Williamsburg, VA 23185, USA\\
        E-mail: \email{mastropaska@gmail.com}}

\author{David G.\ Richards\\
Jefferson Lab, 12000 Jefferson Avenue Suite \# 1, Newport News, VA
23606, USA\\
E-mail: \email{dgr@jlab.org}}

\abstract{We present a calculation of the decay constant of the pion,
  and its lowest-lying three excitations, at three values of the pion
  mass between around 400 and 700 MeV, using anisotropic clover
  lattices.  We use the variational method to determine an optimal
  interpolating operator for each of the states.  We find that the
  decay constant of the first excitation, and more notably of the
  second, is suppressed with respect to that of the ground-state pion,
  but that the suppression shows little dependence on the quark
  mass.}

        \FullConference{31st International Symposium on Lattice Field Theory LATTICE 2013\\
          July 29 - August 3, 2013\\
          Mainz, Germany}

\begin{document}

While obtaining the properties of the ground states of hadrons has
long been a primary goal of lattice QCD, extracting information about
excited hadrons poses numerous challenges. The main difficulty arises
from the faster decay of their Euclidean correlation functions in
comparison with those of the ground states, which leads to the
worsening of the signal-to-noise ratio at increasing temporal
separations. An additional complication arises from the cost of
constructing the necessarily large correlation matrix needed to apply
the variational method, and the implementation of an operator basis
that respects the symmetries of the lattice, yet enables the continuum
quantum numbers to be identified.  Finally, in general we are dealing
with resonances that are unstable under the strong interactions.

Despite all these obstacles, the latest developments in advanced
computational lattice QCD techniques are enabling increasingly precise
quantitative calculations that can both confront existing, and offer
the prospect to predict the outcomes of forthcoming experiments.  Such
experiments are the aim of the 12 GeV upgrade of the Continuous
Electron Beam Accelerator Facility (CEBAF) at Jefferson Lab
 \cite{Dudek:2012vr} with, in particular, a meson spectroscopy program
based on the photoproduction of meson excitations.  The expectation is
that new data produced in this upcoming experiment, combined with
recent lattice QCD results aimed to extracting the spectrum of meson
excited states ~\cite{Dudek:2010wm, Dudek:2011tt}, will represent a
unique opportunity for the study of the nature of confinement
mechanism and for determining the role of the gluonic field in the
hadron spectrum.

The work presented here is also devoted to the study of excited states, but
its emphasis is on the computation of some of their properties. In
general, our goal is to investigate quark distribution amplitudes
which, in case of mesons, can be extracted from the vacuum-to-hadron
matrix elements of quark bilinear operators. These amplitudes would
provide predictions for form factors and transition form factors at
high momentum transfers.  In this talk, we describe
calculations devoted to the study of the leptonic decay properties of
the pion - the lightest system with simple quark-antiquark structure -
and of its excitations.

\section{Pseudoscalar leptonic decay constants}
Charged mesons are allowed to decay, through quark-antiquark
annihilations via a virtual $W$ boson, to a pair of leptons. The decay
width for any pseudo-scalar meson $P$ of a quark content
$q_1~\mbox{and}~\bar{q}_2$ with mass $m_P$ is given as \cite{Beringer:1900zz}
\begin{equation}
\Gamma (P \to l\nu)=\frac{G_f^2}{8\pi}f_P^2m_l^2m_P\left(1-\frac{m_l^2}{m_P^2}\right)|V_{q_1\bar{q}_2}|^2.
\label{eq:20}
\end{equation}
Here $m_l$ is the mass of the lepton $l$, $G_F$ is the Fermi coupling
constant, $V_{q_1\bar{q}_2}$ is the Cabibbo-Kobayashi-Maskawa (CKM)
matrix element between the constituent quarks in $P$, and $f_P$ is the
decay constant related to the wave-function overlap of the quark and
antiquark at the origin. Thus, a charged pion can decay as $\pi \to l
\nu$ (we assume here $\pi ^+\to l^+ \nu_l$ or $\pi^- \to
l^-\bar{\nu}_l$), and its decay constant $f_{\pi}$, which dictates the
strength of these leptonic pion decays, has a significant influence on
many areas of modern physics. Among others pseudo-scalar meson decay
constants, $f_{\pi}$ plays important role for the extracting CKM
matrix elements. Following Eq.~\ref{eq:20}, the leptonic decay width
$\Gamma$ is proportional to $f_P|V_{q_1\bar{q}_2}|$, so theoretically
predicted value of $f_P$ can allow determination of the corresponding
CKM element, which is crucial for testing the flavor sector of the
Standard Model.  As $|V_{ud}|$ has been quite accurately measured in
super-allowed $\beta$-decays, measurements of $\Gamma (\pi^+\to
\mu^+\nu)$ yield a value of $f_{\pi}$.  The most precise determination
of $f_{\pi}$ corresponds to~\cite{Beringer:1900zz}
\begin{equation}
f_{\pi^-}=(130.41\pm 0.03 \pm 0.20) \,\, MeV.
\end{equation}

Lattice QCD provides an \textit{ab initio} means of computing the mass
spectrum and decay constants of pseudo-scalar mesons
nonperturbatively. Thus, for example, recent lattice predictions
\cite{Follana:2007uv, Durr:2010hr, Bazavov:2010hj} for the ratio
$f_K/f_{\pi}$ of $K^-$ and $\pi^-$ decay constants were used in order
to find a value for $|V_{us}|/|V_{ud}|$ which, together with precisely
measured $|V_{ud}|$, provides an independent measure of
$|V_{us}|$. The pion decay constant, determining the strength of $\pi
\pi$ interactions, also serves as an expansion parameter in Chiral
Perturbation Theory (ChiPT) \cite{Weinberg:1978kz,Gasser:1983yg}.
Therefore, reliable lattice QCD determinations of decay constants from
first principles are of fundamental importance.

In this work, we focus on the leptonic decay constants of the excited
states of the pion.  Our motivation is the calculation of the moments
of the quark distribution amplitudes in the pion and its excitations,
for which the decay constants provide the overall
normalization. However, the decay constants of the excitations can
themselves provide insights into QCD-inspired descriptions of
hadrons. The study based on Schwinger-Dyson equations
\cite{Holl:2004fr} predicted significant suppression of the
excited-state pion decay constant with comparison to that one of the
ground state. Similar predictions, based on the QCD-inspired models
and sum rules, also propose remarkably small values for $f_{\pi}^1$;
e.g., \cite{Volkov:1996br} proposed the ratio
$\frac{f_{\pi}^1}{f_{\pi}^0}$ to be of the order of one
percent. There has also been an extraction of the decay constant of
the first excitation of the pion
using lattice QCD \cite{McNeile:2006qy}, in a calculation with two
dynamical quark flavors; this calculation found an increasingly strong
suppression of the decay constant of the first excited state relative
to that of the ground state with decreasing quark mass.
Here we use developments in excited state spectroscopy to compute not
only the decay constant of the ground and first excited state, but of
higher states also.

\section{Computational Method}
Our calculation of the excited-state spectrum of QCD is dependent on
three novel elements: the use of an anisotropic lattice enabling the
time-sliced correlation functions to be examined at small Euclidean
times, the use of the variational method with a large basis of
operators derived from a continuum construction yet which satisfy the
symmetries of the lattice, and an efficient means of computing the
necessary correlation functions through the use of ``distillation''.

The variational approach involves the solution of the generalized
eigenvalue problem
\begin{equation}
C(t)v(t,t_0)=\lambda(t)C(t_0)v(t,t_0),
\label{eq:6}
\end{equation}
where $C(t)$ is an $N \times N$ matrix of correlators constructed from
operators ${\cal O}_i, i = 0\dots N-1$ such that $C_{ij}(t) =
\sum_{\vec{x}}\langle 0 \mid {\cal O}_i(\vec{x},t) {\cal O}^{\dagger}_j(0)
\mid 0 \rangle$.  At sufficiently large $t > t_0$, the ordered
eigenvalues satisfy
\[
\lambda_n(t,t_0) \longrightarrow e^{ - E_n (t - t_0)}
\]
where $E_n$ is the energy of the $n^{\rm th}$ state.  The eigenvalues
are normalized to unity at $t = t_0$, whilst the eigenvectors satisfy
the orthogonality condition $v^{(n')\dagger}C(t_0)v^{(n)}=\delta_{n,
  n'}$, and thereby enable us to construct an \textit{optimal}
operator $\Omega_n$ that couples predominantly to the $n^{\rm th}$
state \cite{Dudek:2009kk}:
\[
\Omega_n = \sqrt{2 E_n} v^{(n)}_i e^{- E_n t_0/2}{\cal O}_i\label{eq:opt}.
\]

The decay constant of the $n^{\rm th}$ excitation of the pion, $\Pi_n$,
is given by the hadron-to-vacuum matrix matrix element of the axial
vector current:
\[
\langle 0 \mid A_{\mu} | \Pi_n> = p_{\mu} f_{\pi}^n.
\]
For a state at rest, we employ the temporal component of the
axial-vector current.  Armed with the optimal interpolating operator
for the $n^{\rm th}$ excited state, we now extract its decay constant
$f_{\pi}^n$ through the two-point correlation function
\begin{equation}
C^n(t) = \sum_{\vec{x}} \langle 0 \mid A_4(\vec{x},t) \Omega^{\dagger}_n(0) \mid 0 \rangle
\longrightarrow e^{- m_n t} m_n f_{\pi}^n, \label{eq:lscorr}
\end{equation}
where now we have identified the energy with the mass of the $n^{\rm
  th}$ excited state $m_n$.

\subsection{Distillation}
Our efficient implementation of the variational method with a large
number of interpolating operators relies on a novel smearing method
denoted by ''distillation'' \cite{Peardon:2009gh}.  Our starting point
is the solution of the Dirac equation
\begin{equation}
\tilde{\tau}^{(k)}_{\alpha\beta}(\vec{x},t;0) =
M^{-1}_{\alpha\beta}(\vec{x},t;0) \xi^{(k)},\label{eq:soln}
\end{equation}
where the $\xi^{(k)}, k = 1 \dots N_{\rm eigen}$ are the lowest
$N_{\rm eigen}$ eigenvectors of a three-dimensional Laplacian at
timeslice $t'=0$.  The
construction of the correlation functions from operators smeared both at
the sink and the source has been described in
detail~\cite{Peardon:2009gh}, but the extension to the calculation of
the smeared-local two-point functions needed here is straightforward:
\begin{eqnarray}
  C_{\mu,i}(t, 0) & = & \frac{1}{V_3} \sum_{\vec{x},\vec{y}}
  \langle 0 \mid A_{\mu} (\vec{x},t) {\cal O}^{\dagger}_i(\vec{y},0) \mid 0 \rangle \nonumber \\
  & = & \frac{1}{V_3} \sum_{\vec{x}} {\rm Tr} [ \gamma_{\mu}
  \tilde{\tau}(\vec{x},t; 0) \Phi_i(0) \gamma_5 \tilde{\tau}(\vec{x}, t;0)^{\dagger}],
\end{eqnarray}
where the trace is over spin, colour and eigenvector indices, $\Phi$
is the representation of the operator ${\cal O}_i$ in terms of the
eigenvectors, and the spatial-volume $V_3$ reflects the time-sliced
sum at the source that is a strength of the distillation method.  The
correlator onto the optimal operator for the $n^{\rm th}$ excited
state immediately follows from Eqn.~\ref{eq:lscorr}.

\section{Calculational Details}
We employ the dynamical $N_f = 2 \oplus 1$ anisotropic lattices
generated by the Hadron Spectrum collaboration, using an anisotropic
``clover'' action with stout smearing in the spatial direction only to
preserve the transfer matrix; the spatial lattice spacing $a_s \simeq
0.12~{\rm fm}$, and the renormalized anisotropy $\xi \equiv a_s/a_t
\simeq 3.5$. Details of the action, the generation of the lattices,
and the setting of the physical scale and quark masses, are contained
in ref.~\cite{Lin:2008pr}.  The parameters of the lattices used here
are shown in Table~\ref{tab:lattices}.
\begin {table}
\centering
\begin{tabular}{cc|ccccc|c}
  $N_s$ & $N_t$ & $a_tm_l$ & $a_tm_s$ & $a_tm_{\pi}$ & $r_0/a_s$& $N_{\rm cfg}$ & $N_{\rm eigen}$ \\ \hline                                       
  16         &   128   &  -0.0743   & -0.0743       &   0.1483(2) & 3.21(1)     &    535 & 64         \\ 
  16         &   128   &  -0.0808   & -0.0743       &  0.0996(6)  & 3.51(1)    &    470  & 64        \\ %\hline
  16         &   128   &  -0.0840   & -0.0743       &  0.0691(6)   & 3.65(1)   &    480  & 64        \\ %\hline
\end{tabular}
\caption{Lattice extents ($N_s$ and $N_t$), the masses of light quark
  $a_tm_l$ and strange quark $a_tm_s$, the pion mass $a_tm_{\pi}$, the
  Sommer scale $r_0$, and the number $N_{\rm cfg}$ of gauge-field
  configurations. The final column shows the number
  $N_{\rm eigen}$ of eigenvectors employed in the implementation of
  ``distillation''~\protect\cite{Peardon:2009gh}.\label{tab:lattices} }
\end{table}
Our operator basis is constructed from the product of quark bilinears,
and of discretisation of the covariant derivative to describe the
spatial structure, namely
\[
\bar{\psi}\Gamma D_{i_0} D_{i_1} \dots
%\overleftrightarrow{D}_i\overleftrightarrow{D}_j\,
\psi,
\]
 which are then projected onto the $A_1^{-+}$ \textit{irrep} of the
cubic group.  We employ up to three derivatives, yielding a basis of
12 operators.

\section{Results}
Our results for the pseudoscalar decay constants using the unimproved
local axial vector current on each of our ensembles are presented in
Table~\ref{tab:decays}.
\begin {table}
\centering
\begin{tabular}{c|cccc}
$m_{\pi}$ (MeV)  & $a_tf_{\pi}^0$ & $a_tf_{\pi}^1$ & $a_tf_{\pi}^2$ & $a_tf_{\pi}^3$ \\ \hline                                       
702          &   0.0551(19)   &  0.0319(67)   & 0.0023(306)      &   0.0320(415)            \\ 
524        &  0.0440(49)  & 0.0254(57)       &  0.0080(338)     &    0.0311(172)          \\ %\hline
 391        &  0.0368(34)   & 0.0207(63)       &  0.0062(115)      &    0.0240(234)          
\end{tabular}
\caption{Unrenormalized values of ground and first three
  excited-state decay constants $|f_{\pi}^N|$ (in lattice units) on
  each of our ensembles.}
\label{tab:decays}
\end{table}
\begin{figure}
\centering
\includegraphics[width=0.8 \textwidth]{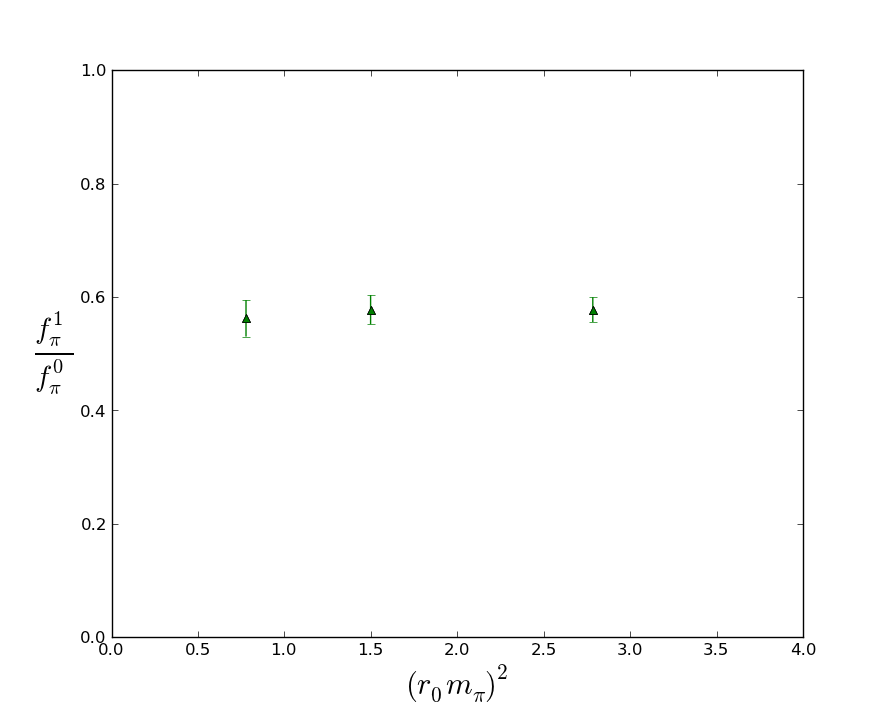}
\caption{The dimensional ratio of the leptonic decay constants
  $f_{\pi}^1$ and $f_\pi^0$ as a function of $m_\pi^2$ in units of the
  Sommer scale $r_0$.
\label{fig:first_decay}}
\end{figure} 
In order to remove dependence both on the lattice spacing, and on the
renormalization constant, we show in Figure~\ref{fig:first_decay} the
ratio of the decay constants of the first and ground states as a
function of the pion mass.  Whilst we see that the first excited state
decay constant is indeed suppressed relative to that of the ground
state, that suppression is quite small and largely independent of the
quark mass over the mass range of this calculation. This is in
contrast to ref.~\cite{McNeile:2006qy}, which shows an increasing
suppression with pion mass over the somewhat similar range of pion
masses used in this calculation, and an even stronger suppression if
an improved axial-vector current is employed; they obtain
$|f_{\pi}^1/f_{\pi}^0|=0.38(11)~{\rm and}~0.078(93)$ for the
unimproved and improved currents respectively in the chiral limit.

\begin{figure}
\centering
\includegraphics[width=0.8 \textwidth]{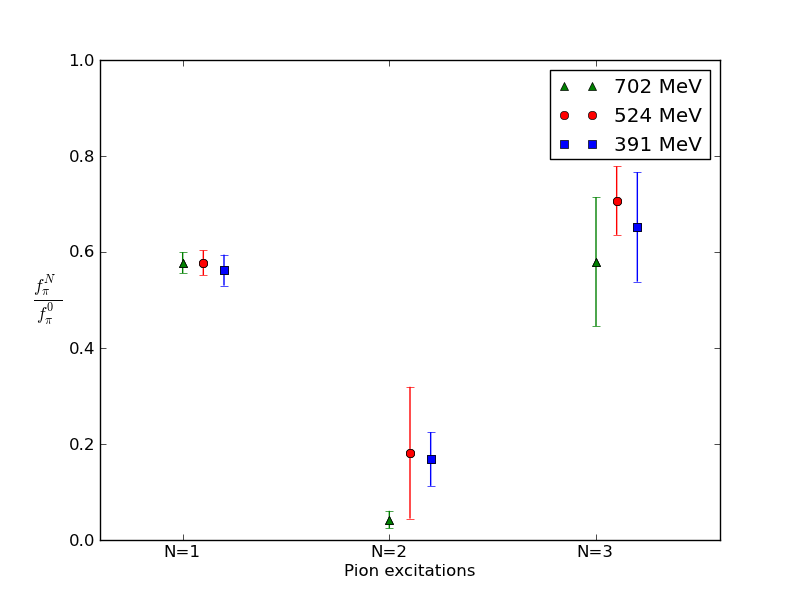}
\caption{Ratios of the excited-state decay constants $f_{\pi}^N$ to the ground-state decay constant $f_{\pi}^0$ for the first 3 pion excitations ($N=1,\, 2, \, 3$).}
\label{fig:all_decays}
\end{figure}

Our ability to isolate the many energy levels in the spectrum enables
us to determine the decay constants not only of the first two states,
but of the second and third excited states.  These are shown for each
of our ensembles in Figure~\ref{fig:all_decays}.  If the decay
constant of the first excited state is only mildly suppressed, we see
that of the second excited state is strongly so, though once again
that suppression does not exhibit a strong dependence on the quark
mass for the unimproved operator over the range of pion masses studied.

\section{Discussion and Conclusions}
In this talk, we undertaken the first steps in investigating the
properties of the excited meson states in QCD by computing the decay
constants both of the pion, and its lowest three excitations.  Our
results show that the optimal operators obtained through the
variational method remain faithful interpolating operators for the
lowest lying excitations in the spectrum when computing
hadron-to-vacuum matrix elements of local operators.  We find that
both the first- and, more notably, the second-excited-state decay
constants are suppressed relative to those of the ground state, but
that this suppression is largely independent of the pion mass.  This
observation differs from that of ref.~\cite{McNeile:2006qy}; our
present calculation does not include operator mixing that they find
largely responsible for the quark mass dependence of the excited-state
decay constants, but note that the anisotropic lattice used in this
calculation has fine spacing in the temporal direction.  Furthermore,
our basis of interpolating operators includes only ``single-hadron''
operators, whose coupling to multihadron decay states is expected to
be suppressed by the volume; we note that the second excited state is
at or above the energy level of the lowest-lying non-interacting
two-meson state on each of our lattices~\cite{Dudek:2010wm}.
Future work will therefore include the inclusion of the operator
improvement term and calculations at larger volumes to expose any
contributions from multihadron states.  Nonetheless, this work clearly
demonstrates that the properties of excited state hadrons are
accessible to lattice calculation.

\acknowledgments
We thank our colleagues within the Hadron Spectrum Collaboration, and
in particular, Jo Dudek, Robert Edwards, Christian Schultz and
Christopher Thomas. We are grateful for discussions with Zak Brown and
Hannes L.L.\ Roberts.  {\tt Chroma}~\cite{Edwards:2004sx} was used to
perform this work on clusters at Jefferson Laboratory under the USQCD
Initiative and the LQCD ARRA project.  We acknowledge support from
U.S. Department of Energy contract DE-AC05-06OR23177, under which
Jefferson Science Associates, LLC, manages and operates Jefferson
Laboratory.

%\begin{thebibliography}{99}
\bibliography{mesondist}
%\end{thebibliography}

\end{document}